\documentclass[preprint,amsmath,amssymb,pre,superscriptaddress]{revtex4-1}
\usepackage{color,longtable}
\usepackage{graphicx}
\usepackage{amssymb,amsfonts,amsmath}
\usepackage{here}

\begin{document}
\title{Liquid-state structural asymmetry governs species-selective crystallization in multicomponent systems}
\author{Rikuya Ishikawa}
\affiliation{%
Department of Physics, Tokyo Metropolitan University, 1-1 Minamioosawa, Hachiouji-shi, Tokyo 192-0397, Japan
}
\author{Kyohei Takae}
\affiliation{%
Department of Mechanical and Physical Engineering, Tottori University, 4-101 Koyama-cho Minami, Tottori 680-8550, Japan
}%
\author{Daisuke Takegami}
\affiliation{%
Department of Physics, Tokyo Metropolitan University, 1-1 Minamioosawa, Hachiouji-shi, Tokyo 192-0397, Japan
}%
\affiliation{Max Planck Institute for Chemical Physics of Solids, N{\"o}thnitzer Stra{\ss}e 40, 01187 Dresden, Germany}
\author{Yoshikazu Mizuguchi}
\affiliation{%
Department of Physics, Tokyo Metropolitan University, 1-1 Minamioosawa, Hachiouji-shi, Tokyo 192-0397, Japan
}%
\author{Rei Kurita}
\affiliation{%
Department of Physics, Tokyo Metropolitan University, 1-1 Minamioosawa, Hachiouji-shi, Tokyo 192-0397, Japan
}%
\date{\today}
\begin{abstract}
Multicomponent crystals are often assumed to form nearly random solid solutions when thermodynamically stable. 
However, crystal growth proceeds from structurally heterogeneous liquids, raising the possibility that the liquid state may influence which species are incorporated into the growing crystal. 
Here we demonstrate that liquid-state structural asymmetry can induce species-selective crystallization in multicomponent systems. 
Using molecular dynamics simulations of a multivalent rocksalt-type model (AgPbBiTe$_3$), we find that cations with higher valence readily form locally crystal-compatible coordination environments in the liquid and are efficiently incorporated into the growing lattice, whereas lower-valence cations exhibit more disordered liquid coordination and attach less efficiently at the crystal-liquid interface. 
This asymmetry leads to species-selective incorporation and slower crystal growth. 
Depth-resolved photoelectron spectroscopy measurements on AgPbBiTe$_3$ further reveal enhanced Ag concentration near grain-boundary and surface regions, consistent with the selective incorporation predicted by the simulations. 
These results demonstrate that structural compatibility between liquid-state structure and the target crystal motif governs selective incorporation during crystallization, providing a general kinetic mechanism by which compositional heterogeneity can emerge during growth of multicomponent crystals.
\end{abstract}


\maketitle

\section{Introduction}
The formation of a random solid solution is commonly assumed when a multicomponent phase is thermodynamically stable~\cite{Yeh2004,Cantor2004}. In single-component systems, liquid-state structure is known to influence crystallization pathways and nucleation kinetics~\cite{Oxtoby}. For example, local ordering in water promotes ice formation~\cite{Russo2012}, and density fluctuations near liquid-liquid critical points accelerate crystallization in molecular liquids and protein solutions~\cite{Kurita2019,DFrenkel1997,Galkin1}. In such cases, however, all particles are identical and the liquid does not discriminate between species. In multicomponent liquids, by contrast, different species experience distinct local coordination environments and interaction strengths. If these liquid-state asymmetries persist at the crystal--liquid interface, incorporation need not be equivalent among species. Under such conditions, the assumption of compositional randomness in multicomponent crystals need not hold.

Indeed, deviations from perfect randomness are widely observed in multicomponent solids. Short-range order and mesoscale compositional correlations have been reported in high-entropy alloys and related systems~\cite{Cowley1950,Zhang2017,George2019}, and such heterogeneity can strongly influence physical properties. For example, in multicomponent ionic conductors, the percolation of mobile ions within a disordered lattice critically determines macroscopic transport behavior~\cite{Ishikawa2025}. These observations demonstrate that internal atomic arrangements can directly govern macroscopic properties.

The formation pathway of such compositional heterogeneity, however, remains unclear. In many complex alloys and ionic crystals, atomic diffusion is extremely slow, and chemical equilibration over mesoscopic length scales can far exceed typical solidification or processing times. If extensive rearrangement in the solid state is kinetically hindered, it becomes difficult to attribute observed compositional correlations solely to post-solidification equilibration~\cite{Ishikawa2024}. This suggests that at least part of the heterogeneity may be established during crystal growth itself.

\begin{figure}[htbp]
\centering
\includegraphics[width=14cm]{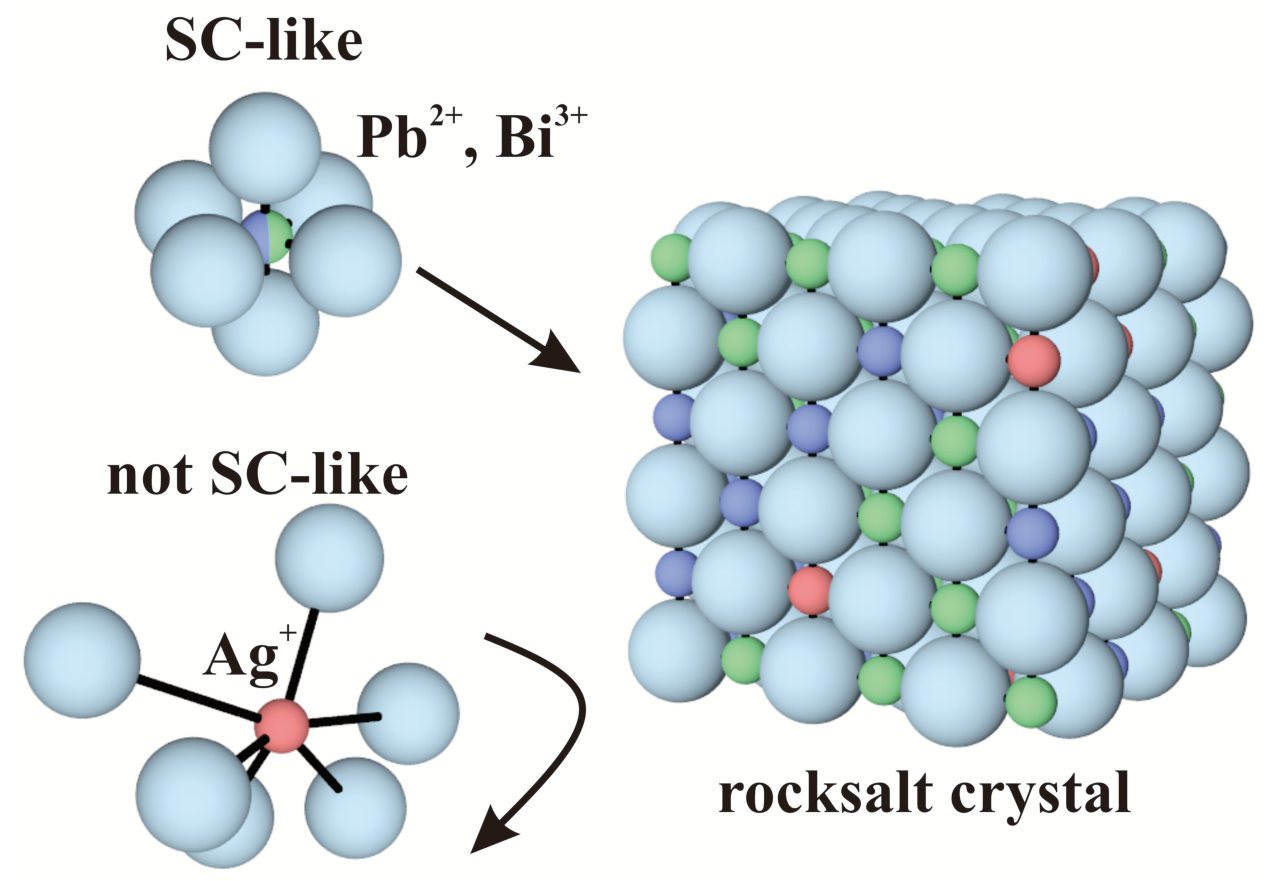}
\caption{
Schematic illustration of charge-selective crystallization.
Higher-valence cations form locally rocksalt-compatible coordination with Te$^{2-}$ and attach efficiently to the growth front.
In contrast, Ag$^+$ exhibits more disordered liquid coordination and is incorporated less efficiently.
This liquid-state structural asymmetry leads to selective incorporation and compositional bias during crystal growth.
}
\label{GA}
\end{figure}

To examine whether crystal growth can directly select composition, we employ a minimal yet physically representative multivalent model that enables large-scale and long-time simulations of crystallization dynamics. Using molecular dynamics simulations, we directly follow crystal growth from the liquid and find that the resulting crystals exhibit pronounced compositional inhomogeneity: cations with smaller positive charge are incorporated far less efficiently into the lattice and remain enriched in the surrounding liquid (Fig.~\ref{GA}). By comparison with a charge-unified reference system, we demonstrate that this bias does not originate from geometric size disparity but from species-dependent coordination environments already present in the liquid. These liquid-state structural differences selectively influence attachment kinetics at the crystal--liquid interface, producing species-selective incorporation during growth. Our results therefore identify a microscopic mechanism in which liquid-state structural asymmetry controls compositional bias during crystallization. Because differences in local coordination are a generic feature of multicomponent liquids, this mechanism is not specific to ionic systems but may operate broadly in alloys, complex oxides, and other compositionally complex materials. More generally, whenever different species exhibit distinct liquid-state structural compatibility with the emerging crystal motif, crystal growth becomes inherently selective.

\section{Results}
\subsection{Species-selective crystallization during crystal growth}
We investigated crystallization in a minimal yet physically representative multivalent ionic model,
AgPbBiTe$_3$, in which particles interact through excluded-volume repulsion and Coulomb forces.
This rocksalt-type system can be synthesized experimentally, allowing direct comparison with measured properties~\cite{Mizuguchi2023}.
In particular, the simulated melting temperature and sound velocity are in reasonable agreement with experimental values, indicating that the model captures the essential physics of multicomponent telluride crystals~\cite{Ishikawa2024}.
In AgPbBiTe$_3$, the cations carry distinct valences (Ag$^{+}$, Pb$^{2+}$, and Bi$^{3+}$), resulting in intrinsic charge dispersity.
As a reference, we also simulated a charge-unified ideal system (CUI), in which the ionic radii of the cations are unchanged but all cation charges are set to $+2$.
This construction isolates the role of charge diversity while preserving geometric size effects.
Starting from an equilibrated liquid at $T = 4.4$, the systems were rapidly quenched below their melting temperatures.
Within the simulation time window, crystallization occurs at $T = 2.0$ for AgPbBiTe$_3$ and at $T = 2.4$ for the CUI system, reflecting the higher melting temperature of the latter.
Details of the simulation protocol are provided in Methods, and the corresponding volume evolution and radial distribution functions are shown in Supplementary Information.

\begin{figure}[htbp]
\centering
\includegraphics[width=14cm]{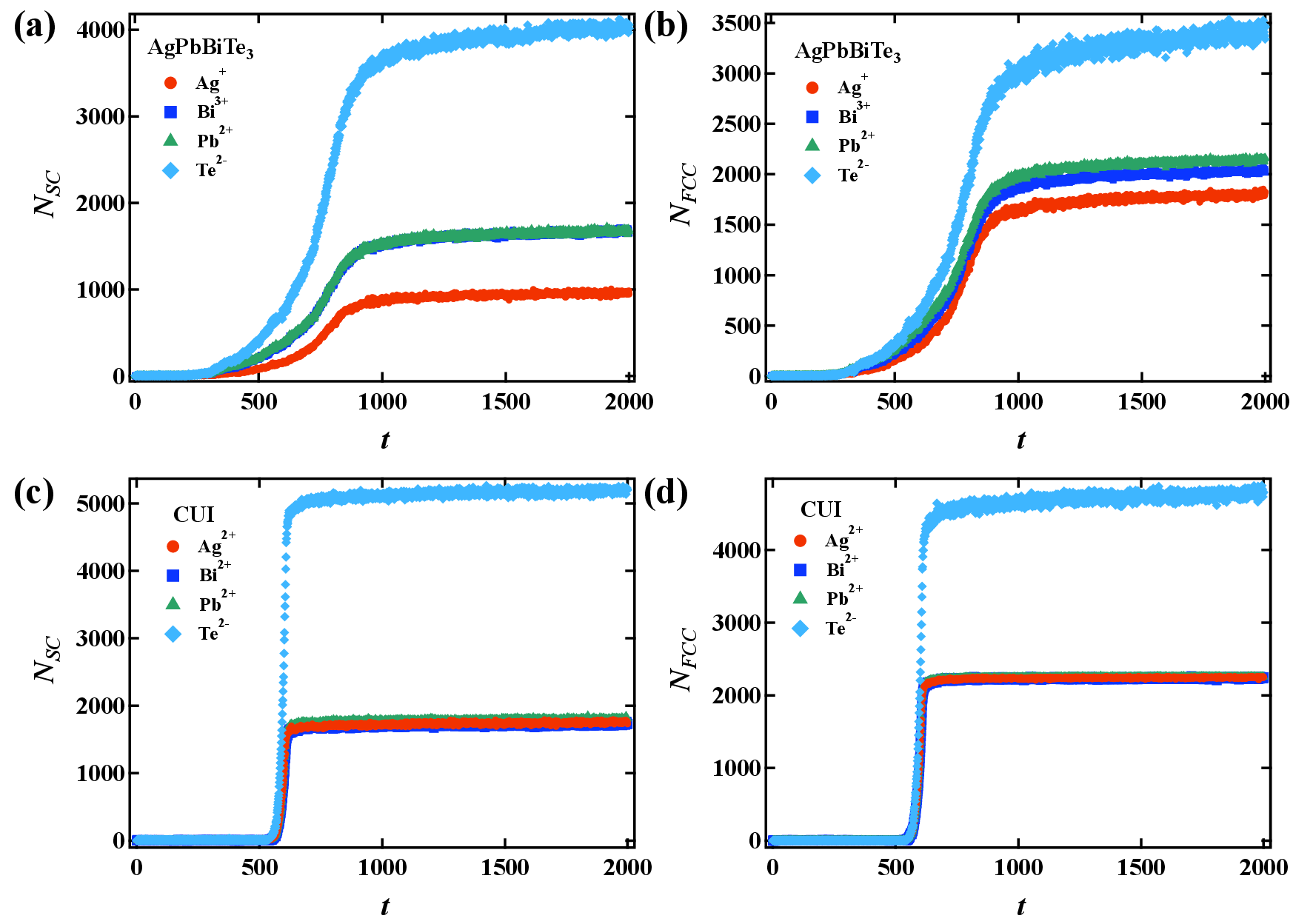}
\caption{Time evolution of the number of ions belonging to the largest crystalline cluster for AgPbBiTe$_3$ and the charge-unified ideal (CUI) system.
Panels (a) and (b) show results for AgPbBiTe$_3$ using the SC and FCC criteria, respectively, while panels (c) and (d) show the corresponding results for CUI.
AgPbBiTe$_3$ exhibits a markedly reduced incorporation of Ag$^+$, resulting in non-uniform crystallization, whereas CUI shows uniform incorporation across cation species.
These differences demonstrate that the compositional inhomogeneity arises from charge diversity rather than ionic size.
}
\label{number}
\end{figure}

To quantify the compositional uniformity during crystal growth, we monitored the number of ions incorporated into the largest crystalline cluster as a function of time.
Figure~\ref{number}(a) shows the evolution of incorporated ion counts in AgPbBiTe$_3$ using the simple-cubic (SC) criterion.
While Pb$^{2+}$ and Bi$^{3+}$ are incorporated in nearly equal amounts, Ag$^{+}$ is markedly underrepresented throughout growth.
This deficiency persists over the entire crystallization process, indicating suppressed incorporation of Ag$^{+}$ relative to the higher-valence cations. 
To confirm that this imbalance is not an artifact of the classification scheme, we repeated the analysis using an alternative face-centred cubic (FCC) criterion, which differs primarily in its treatment of interfacial and grain-boundary regions. 
The reduced Ag content remains robust under this definition (Fig.~\ref{number}(b)),
demonstrating that the compositional bias is not sensitive to the detection method.

In contrast, the charge-unified ideal system (CUI) exhibits nearly identical incorporation dynamics for all cation species (Fig.~\ref{number}(c, d)).
Because the CUI preserves ionic size but removes charge dispersity, this comparison shows that the suppressed incorporation of Ag$^{+}$ originates from valence differences rather than geometric effects.
In addition to the compositional bias, the crystallization dynamics in AgPbBiTe$_3$ are noticeably broader in time compared to the CUI system, indicating slower crystal growth.
We return to this kinetic aspect below.

\begin{figure}[htbp]
\centering
\includegraphics[width=14cm]{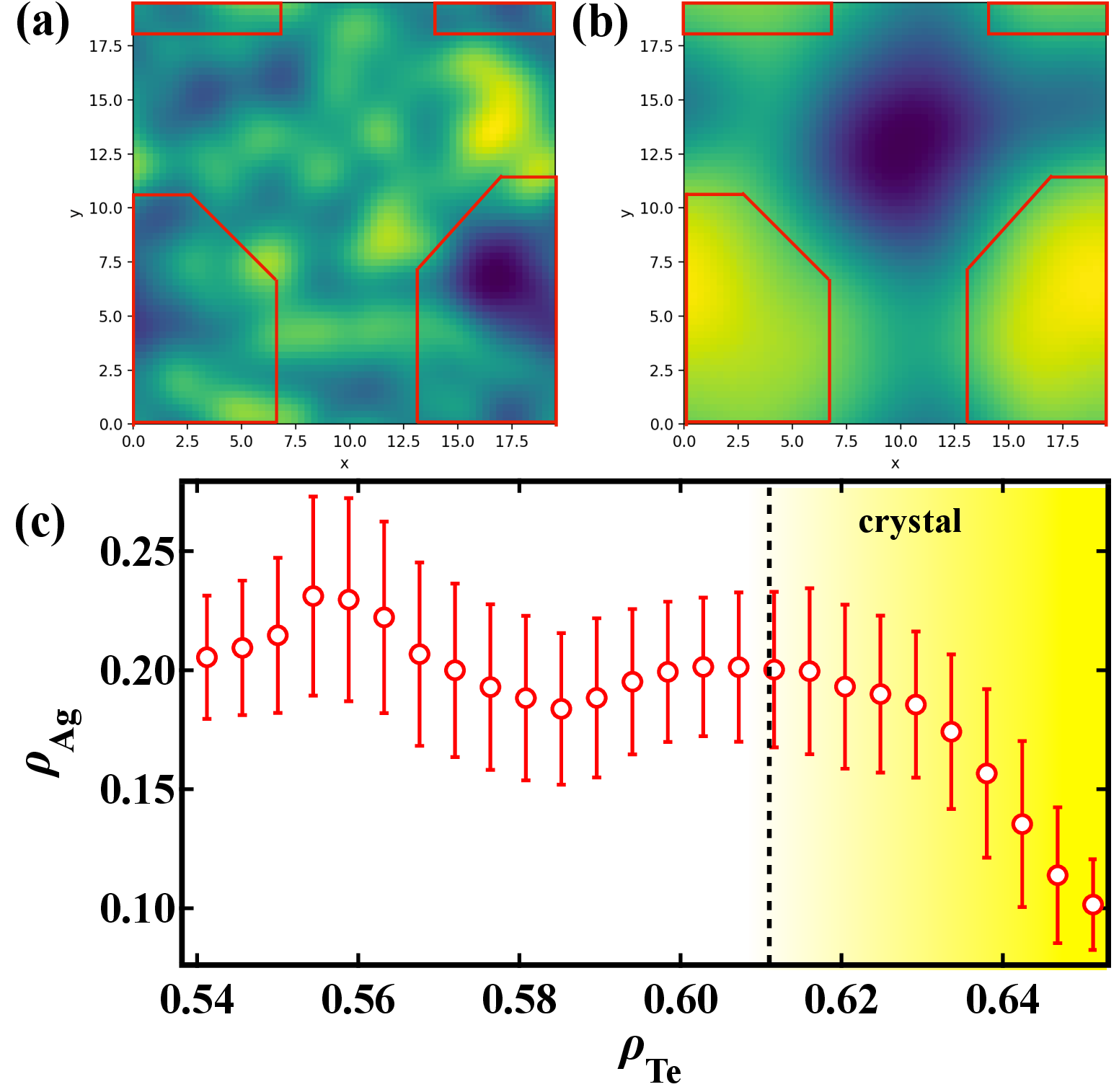}
\caption{Color plot of (a) Ag$^{+}$ density and (b) Te$^{2-}$ density in the $xy$ plane at $z = 10$ and $t = 800$ (yellow: high density, blue: low density).
The region enclosed by the red line corresponds to the high-density Te$^{2-}$ domain, identifying the crystalline region.
Within this region the Ag$^{+}$ density is depleted, while enhanced Ag$^{+}$ density is observed near the surrounding growth front.
(c) Correlation between $\rho_{Ag}$ and $\rho_{Te}$.
Regions with $\rho_{Te} > 0.61$ correspond to crystalline order, where $\rho_{Ag}$ is reduced, consistent with the suppressed incorporation of Ag$^{+}$ ions in the crystal.
}
\label{smearing}
\end{figure}

To visualize the spatial distribution of cations during crystal growth,
we constructed coarse-grained density fields by Gaussian smearing of particle positions.
Figure~\ref{smearing}(a) and (b) show the Ag$^{+}$ density in the $xy$ plane at $z = 10$ and $t = 800$.
For reference, the corresponding Te$^{2-}$ density at the same location is also shown.
Because crystalline regions exhibit higher density than the surrounding liquid,
the area of elevated Te$^{2-}$ density identifies the growing crystal (enclosed region).
In contrast, the Ag$^{+}$ density is depleted inside this region.
Figure~\ref{smearing}(c) shows the correlation between $\rho_{Ag}$ and $\rho_{Te}$.
Regions with $\rho_{Te} > 0.61$ correspond to crystalline order, and in this region $\rho_{Ag}$ is reduced,
which is consistent with the lower abundance of Ag$^{+}$ ions within the crystal.
In contrast, the smearing images show that Ag$^{+}$ density is enhanced in the surrounding region.
Thus, Ag$^{+}$ ions are scarce within the crystalline core but accumulate near the crystal--liquid interface.
Quantitatively, the interfacial fraction of Ag$^{+}$ reaches nearly $0.5$,
substantially exceeding the random expectation of $1/3$ (see Supplementary Information for details of the analysis).
This spatial segregation is fully consistent with the suppressed incorporation of Ag$^{+}$ observed in Fig.~\ref{number}.

\subsection{Densification pathway and growth kinetics}
\begin{figure}[htbp]
\centering
\includegraphics[width=14cm]{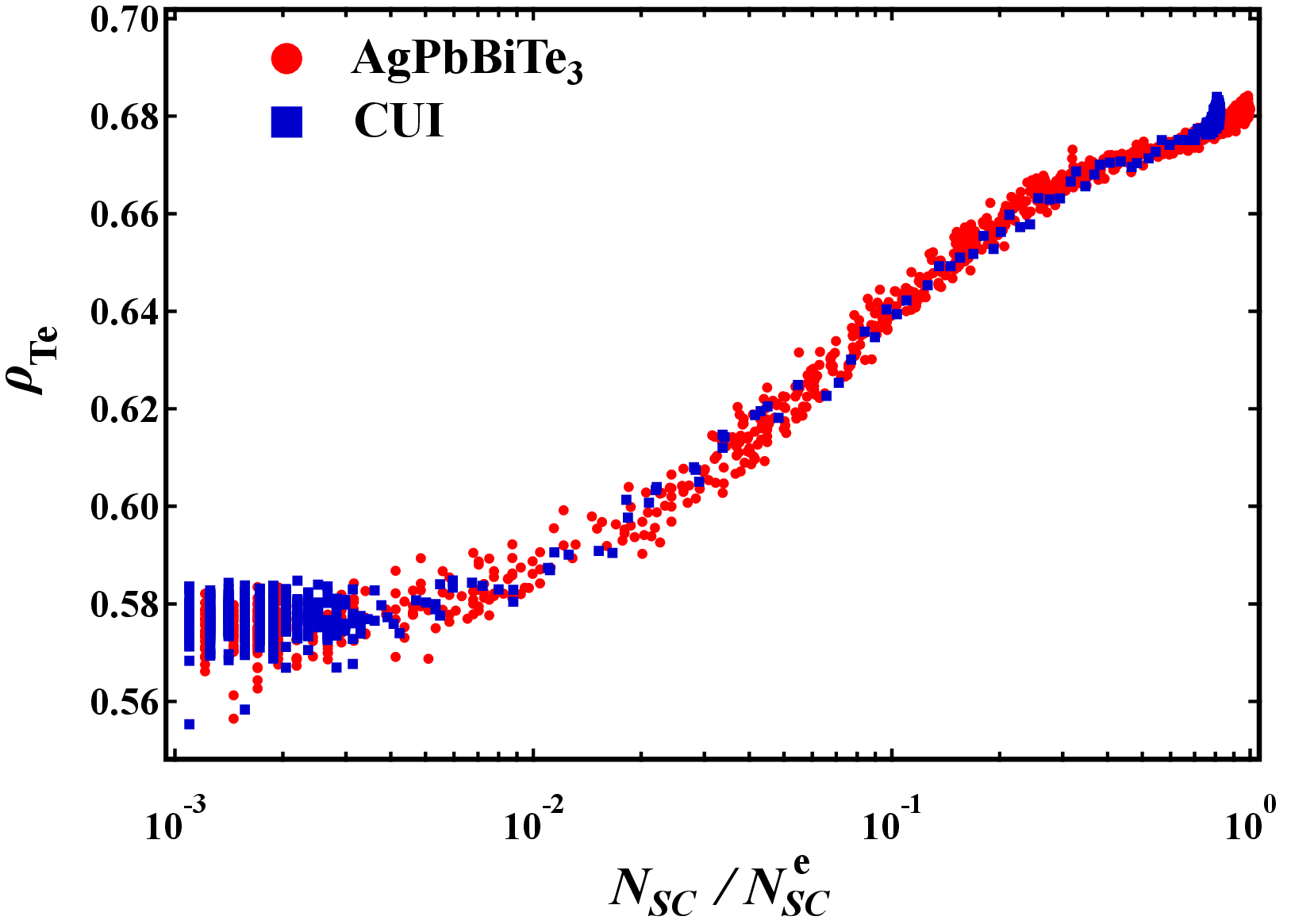}
\caption{\textbf{Thermodynamic densification pathway during crystallization.}
Average local density of Te$^{2-}$ as a function of the crystallization progress
($N_{SC}/N_{SC}^{e}$, where $N_{SC}^{e}$ is the number of crystalline particles at the end of the simulation)
for AgPbBiTe$_3$ and the CUI system.
The nearly identical curves indicate that the thermodynamic pathway of densification is essentially unchanged between the two systems.
}
\label{density2}
\end{figure}

To determine whether the slower growth in AgPbBiTe$_3$ originates from thermodynamic or kinetic effects, we examined the relationship between crystal size and density during crystallization. 
This densification pathway is known to correspond to the time evolution along the free-energy ridge during crystal growth~\cite{Oxtoby}. 
The local density of particle $i$ was computed from its Voronoi volume and ionic diameter. 
Figure~\ref{density2} shows the average local density as a function of crystalline cluster size for AgPbBiTe$_3$ and the charge-unified ideal (CUI) system. 
The nearly identical curves indicate that the thermodynamic densification pathway during crystallization is essentially unchanged by charge unification. 
Therefore, the slower growth observed in AgPbBiTe$_3$ cannot be attributed to differences in thermodynamic driving forces. 
Instead, the growth asymmetry must originate from kinetic processes at the crystal--liquid interface. 
Consistently, the crystalline domain in AgPbBiTe$_3$ expands significantly more slowly than in the CUI system (Fig.~\ref{number}). 
The suppressed incorporation of Ag$^{+}$ therefore reflects a charge-dependent kinetic bottleneck at the growth front.

\subsection{Liquid-state structural origin of species-selective growth}
\begin{figure}[htbp]
\centering
\includegraphics[width=14cm]{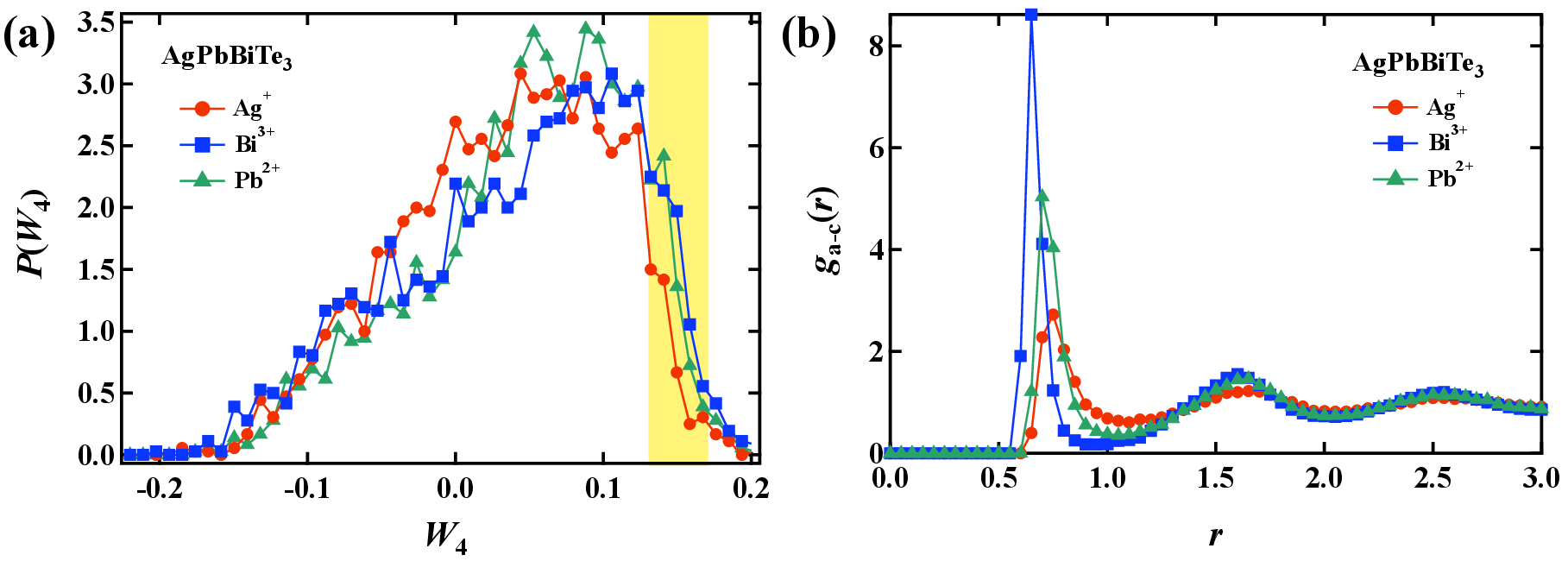}
\caption{(a) $\hat W_4$ histograms for each cation at $t$ = 200 in AgPbBiTe$_3$. In the multivalent system, the weaker Coulomb attraction of Ag$^+$ suppresses the formation of locally SC-like environments in the liquid, indicating that the structural asymmetry underlying charge-selective incorporation is already present before crystallization. The yellow region in the figure indicates the area classified as the SC-like structure.   
(b) Te-cation radial distribution functions at $t$ = 200 in AgPbBiTe$_3$. The broader and lower first peak for Ag$^+$ in AgPbBiTe$_3$ indicates larger and more widely distributed Ag-Te neighbor distances, consistent with the reduced local density. These liquid-state structural differences account for the inefficient incorporation of Ag$^+$ at the crystal-liquid interface.
}
\label{liquid}
\end{figure}

To identify the microscopic origin of the charge-selective incorporation,
we examined the structure of the liquid prior to nucleation.
Figures~\ref{liquid}(a) show the distributions of the fourth-order bond-orientational order parameter $\hat W_4$
at $t = 200$, when no crystalline nuclei are present.
Large positive values of $\hat W_4$ correspond to local environments resembling simple-cubic (SC) symmetry. 
The yellow region in the figure indicates the area classified as the SC crystal.
In AgPbBiTe$_3$, Ag$^{+}$ exhibits a suppressed population at large $\hat W_4$,
whereas Pb$^{2+}$ and Bi$^{3+}$ display nearly identical distributions.
These results demonstrate that in the multivalent system, Ag$^{+}$ is less likely to form locally SC-like coordination environments already in the liquid state.

This structural asymmetry is further reflected in the Te--cation radial distribution functions $g(r)$ (Figs.~\ref{liquid}(b)).
In AgPbBiTe$_3$, the first peak of $g(r)$ depends strongly on cation valence:
higher-valence cations exhibit sharper and higher peaks, indicating stronger binding to Te$^{2-}$.
The broader and lower first peak for Ag$^{+}$ implies larger and more widely distributed Ag--Te separations,
consistent with the reduced local density around Ag$^{+}$ observed at the early stage of crystallization.

Taken together, these results indicate that the liquid already contains a charge-dependent structural heterogeneity. 
Since crystal growth from a liquid requires local structural rearrangement prior to incorporation, differences in the liquid-state environment can generate an effective attachment barrier at the crystal-liquid interface~\cite{Russo2012,Sosso2016}. 
In the present system, cations whose liquid environments more closely resemble the crystalline motif attach more readily at the growth front, whereas Ag$^{+}$ experiences a larger kinetic barrier due to its more disordered coordination. 
The charge-selective growth, spatial Ag accumulation at the interface, and the overall slowdown of crystallization thus originate from structural asymmetry encoded in the liquid phase.
Experimentally, crystallization in multicomponent systems has often been reported to proceed more slowly than in simpler compositions~\cite{Lee2026}.
The observed species-selective growth and slower crystallization therefore originate from structural asymmetry encoded in the liquid phase.

\subsection{Kinetic trapping of compositional bias}
\begin{figure}[htbp]
\centering
\includegraphics[width=14cm]{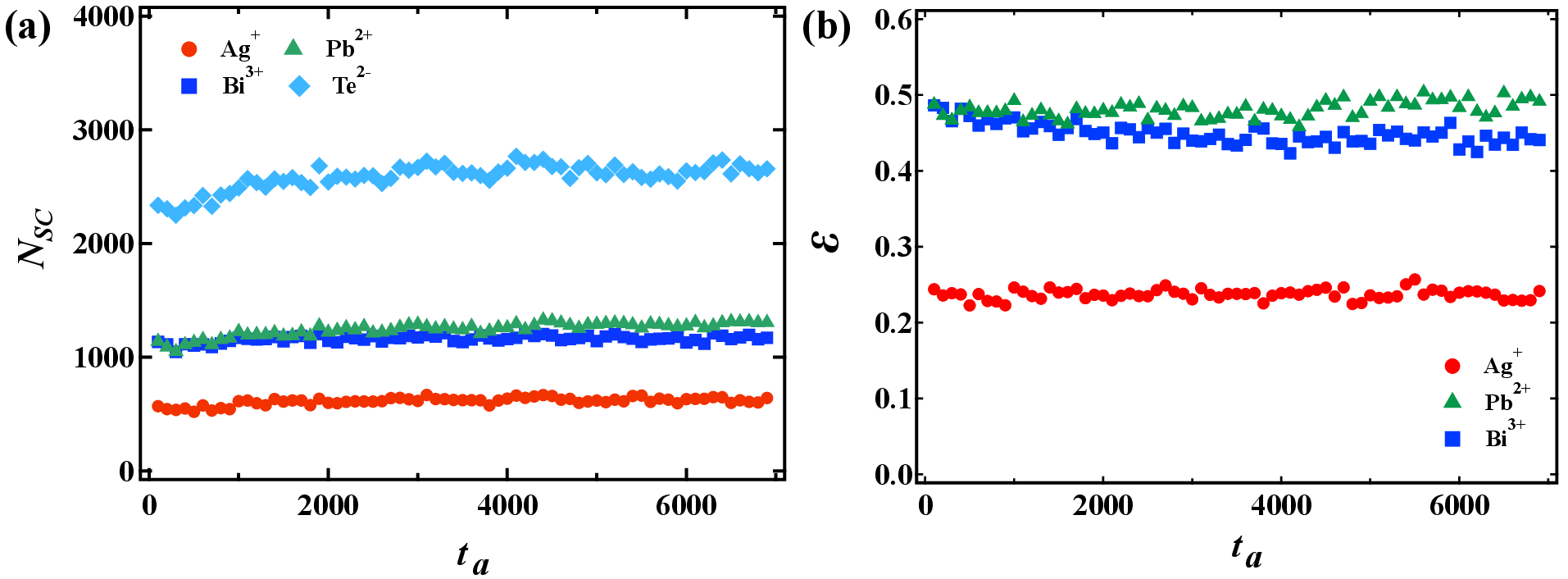}
\caption{
(a) Time evolution of the number of each ion species in the crystalline domain during annealing, and 
(b) time evolution of the cation-to-Te ratios after heating a crystal formed at $T = 2.0$ to $T = 2.61$ ($T_m - T = 0.01$). 
Although the crystalline domain grows slightly during annealing, the Ag$^+$ fraction remains low and shows no significant relaxation.
}
\label{anneal}
\end{figure}

To examine whether the compositional bias relaxes near equilibrium,
we performed additional simulations in which the crystal formed at $T = 2.0$
was heated to temperatures just below the melting point
($T_m - T = 0.01$) and annealed for extended times shown as in Fig.~\ref{anneal}(a).  
Immediately after heating, grain boundaries partially melt,
leading to a decrease in the average density.
During annealing, slow crystallization proceeds gradually,
particularly near grain-boundary regions,
indicating that structural rearrangement remains possible close to the melting temperature.
Despite this continued structural evolution,
the cation composition within the crystalline domain shows no measurable change.
Figure \ref{anneal}(b) shows the Ag/Pb/Bi ratios to Te remain essentially constant over the annealing period.
Although the accessible simulation time and system size are finite,
these results indicate that the compositional bias established during growth does not readily relax even under near melting temperature.

\subsection{Experimental verification}
\begin{figure}[htbp]
\centering
\includegraphics[width=14cm]{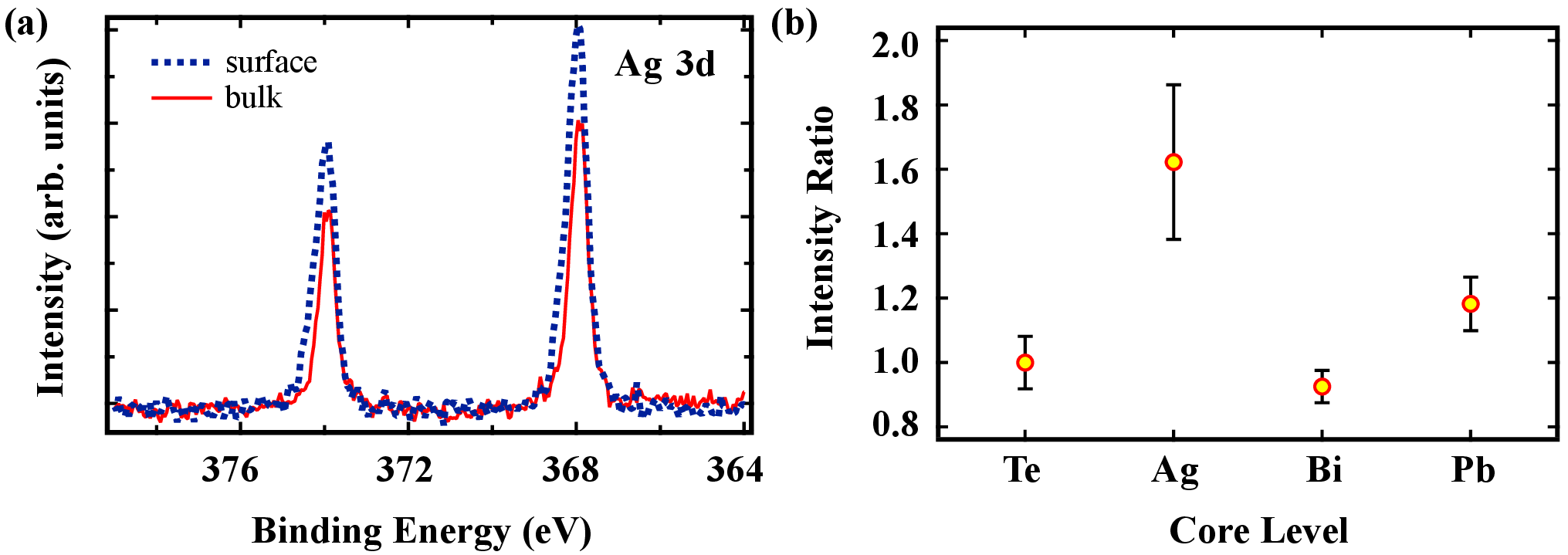}
\caption{Photoemission core level spectra of surface-sensitive spectra (dotted line) and blulk-sensitive spectra (solid line), after a standard Shirley background subtraction and normalized to the total integrated intensity of the Te~$4d$ spectra. 
We observe that the intensity of the Ag significantly increases for its surface-sensitive spectra.
(b) The relative ratios obtained from the integrated intensities of each of the sub-shell peaks. The highest increase on the Ag is observed, which is consistent with the theoretically expected behaviour.
}
\label{ratio2}
\end{figure}

To examine whether the behavior observed in the simulations also occurs in real materials, 
we performed depth resolved X-ray photoelectron spectroscopy (XPS) measurements on AgPbBiTe$_3$ samples. 
By changing the incidence angle, surface-sensitive and bulk-sensitive spectra can be compared. 
Experimental details are described in the SI. 
Figure~\ref{ratio2}(a) shows the Ag$^+$ spectrum normalized by the Te$^{2-}$ intensity. 
The solid line represents the surface-sensitive spectrum, while the dotted line corresponds to the bulk-sensitive spectrum. 
Spectra for the other elements are shown in Fig.~\ref{XPS} in the SI. 
Figure~\ref{ratio2}(b) shows the relative ratios obtained from the integrated intensities of the sub-shell peaks. 
Among the cations, the largest increase is observed for Ag, followed by Pb and then Bi, consistent with the behavior expected from the simulations. 
We note that the quantitative values depend on the normalization procedure, here taken with respect to the Te intensity, but the qualitative trend is robust. 
We also note that the enhanced Ag concentration near grain-boundary regions persists even several days after synthesis. 
This persistence of compositional bias is consistent with the kinetic trapping observed in our simulations.

\section{Discussion}
Whether such growth-induced compositional imbalance ultimately relaxes is determined by the competition between selective incorporation at the crystal--liquid interface and subsequent ionic diffusion within the solid. 
If solid-state diffusion is sufficiently fast, local imbalance generated during growth can relax over time, yielding an apparently random solid solution at long times. 
In contrast, when diffusion is slow on experimentally relevant timescales, the bias becomes kinetically frozen. 
Our annealing simulations near the melting temperature show that, although gradual structural relaxation and grain-boundary crystallization remain possible, the overall cation ratios do not recover toward uniformity. 
These findings complement thermodynamic arguments for disorder stabilization in compositionally complex materials~\cite{Bianchini2020}. 
While configurational entropy may favor random solid solutions at equilibrium, the present results demonstrate that kinetic constraints during growth can generate persistent compositional heterogeneity. 
Thus, achieving compositional randomness in multicomponent crystals requires consideration of both equilibrium stability and diffusion-limited growth dynamics.

The present results provide a microscopic mechanism for the compositional heterogeneity observed both in the simulations and in experiment. 
Charge-selective incorporation during crystal growth produces Ag-deficient crystalline interiors and interfacial Ag accumulation, consistent with the enhanced Ag concentration detected near grain-boundary regions in AgPbBiTe$_3$. 
Grain-boundary segregation and compositional heterogeneity have been reported in a wide range of multicomponent materials, including high-entropy alloys~\cite{Zhang2017, Bianchini2020}, complex oxides, and hydrogen-storage compounds~\cite{George2019}, and are often attributed to thermodynamic stabilization or slow diffusion in the solid state. 
The present results demonstrate that such heterogeneity can also originate during crystal growth itself, as a consequence of liquid-state structural asymmetry that selectively controls incorporation at the crystal-liquid interface. 
More generally, whenever different species exhibit distinct structural compatibility with the target lattice, crystal growth becomes selective, providing a general kinetic route to compositional bias in complex materials.

\section{Conclusion}
In this study, we have shown that crystallization in multicomponent ionic systems can intrinsically generate compositional heterogeneity through charge-selective incorporation at the crystal--liquid interface.
Using molecular dynamics simulations of AgPbBiTe$_3$, we demonstrated that Ag$^{+}$ is incorporated significantly less efficiently than higher-valence cations, leading to both slowed growth and a persistent Ag deficiency in the solid.
Comparison with a charge-unified reference system establishes that this effect arises from valence diversity rather than geometric factors.
Our analysis further reveals that the origin of this selective growth is encoded in the liquid state:
differences in local coordination and structural compatibility with the rocksalt lattice determine incorporation efficiency.
Whether the resulting compositional bias relaxes depends on the competition between interfacial selection and solid-state diffusion.
When diffusion is slow on experimental timescales, as in the present system, the bias becomes kinetically frozen.

These results demonstrate that compositional uniformity in multicomponent ionic crystals is governed not only by thermodynamic stability, but by how liquid-state local structures map onto the coordination motifs of the emerging crystal.
Whenever different species exhibit distinct liquid-phase structural compatibility with the target lattice, growth becomes inherently selective. 
Thus, liquid-state structural heterogeneity provides a general microscopic origin for diffusion-limited compositional bias in complex crystalline materials.

\section*{Methods}
\subsection*{Molecular dynamics simulations}
We performed molecular dynamics (MD) simulations of a multicomponent ionic system representing rocksalt-type AgPbBiTe$_3$.
Particles interacted via a Weeks-Chandler-Andersen (WCA) repulsive potential combined with Coulomb interactions.
The equations of motion were integrated using the leap-frog scheme with periodic boundary conditions applied in all directions.
Long-range electrostatics were computed using the Ewald summation method~\cite{Allen}.
The unit of length and mass were normalized by the ionic diameter and mass of Te$^{2-}$ ($d_{\rm Te} = 4.42~\mathrm{\AA}$~\cite{Shannon1976}, $m = 127.6~\mathrm{g\,mol^{-1}}$~\cite{Thomas2022}), respectively.
To isolate the role of charge dispersity, we additionally simulated a charge-unified ideal (CUI) system in which all cations were assigned valence +2 while retaining their original ionic radii.
All simulations were performed at constant pressure using an Andersen barostat and a Nose-Hoover thermostat.
After equilibration in the liquid phase above the melting temperature, the systems were rapidly quenched and crystallization dynamics were monitored.
The system size was $N = 24^3$ particles.
The melting temperature was determined by gradually heating a crystalline configuration until structural order was lost.
Further details of interaction parameters, reduced units, and Ewald settings are provided in the Supplementary Information.

\subsection*{Structural analysis}
Crystalline particles were identified using bond-orientational order parameters~\cite{Steinhardt1983, Kurita2010}.
Because the stable phase adopts the rocksalt structure, two complementary criteria were employed:
(i) a simple-cubic (SC) criterion applied to all particles, and
(ii) an FCC-based criterion applied to the Te$^{2-}$ sublattice.
Both involve combined thresholds of fourth-, sixth-, and eighth-order orientational symmetry components.
These independent measures consistently detected nucleation and crystal growth.
Full definitions of order parameters and threshold values are provided in the Supplementary Information.

\subsection*{Experimental characterization}
Polycrystalline AgPbBiTe$_3$ samples were synthesized following established procedures~\cite{sawahara2023}.
The samples were mechanically fractured prior to measurement, exposing internal grain-boundary surfaces.
Depth resolved X-ray photoelectron spectroscopy (XPS) measurements were performed to probe local compositional variations at and near grain-boundary regions by changing the angle between the incident light and the detector~\cite{Takegami2019}.
Elemental distributions were analyzed from the exposed surfaces.
Additional experimental details and representative spectra are provided in the Supplementary Information.

\section*{Acknowledgements}
We thank A. Seshita for assistance with the experiments. 
R. I. was supported by the MIYAKO-MIRAI Project of Tokyo Metropolitan University and JSPS Research Fellow Grant Number 24KJ1854. 
K. T. was supported by JSPS KAKENHI Grant Number 24K00594 and 25H01978. 
Y. M. was supported by TMU research funds for young scientists. 

\section*{AUTHORS CONTRIBUTIONS}
R.~K. conceived the project. R.~I. and K. T. performed the numerical simulations and analyzed the data. D. Takegami performed XPS experiments. Y. Mizuguchi prepared the samples for the experiments.R.~I. and R.~K. wrote the manuscript.

\section*{COMPETING INTERESTS STATEMENT}
The authors declare that they have no competing interests. 

\section*{CORRESPONDENCE}
Correspondence and requests for materials should be addressed to R.~K. (kurita@tmu.ac.jp).

\section*{Availability of Data and Materials}
All data generated or analyzed during this study are included in this published article and its supplementary information files.

\bibliography{crystal}

\newpage
\section*{Supporting information}
\setcounter{figure}{0}
\renewcommand{\thefigure}{S\arabic{figure}}

\section*{Supplementary Methods}

\subsection*{Interaction potentials and units}

Particles $i$ and $j$ interact via a Weeks-Chandler-Andersen (WCA) potential
\[
U_{\rm WCA}(r_{ij}) =
\begin{cases}
4\epsilon \left[ \left(\frac{d_{ij}}{r_{ij}}\right)^{12}
- \left(\frac{d_{ij}}{r_{ij}}\right)^6 \right] + \epsilon,
& r_{ij} < 2^{1/6} d_{ij} \\
0, & r_{ij} \ge 2^{1/6} d_{ij}
\end{cases}
\]
combined with Coulomb interactions
\[
U_q(r_{ij}) = -\frac{k q_i q_j}{r_{ij}}.
\]
The equation of motion is
\[
m \frac{d^2 \mathbf{r}_i}{dt^2}
= -\sum_{j \ne i} \nabla \left( U_{\rm WCA} + U_q \right).
\]
The Te$^{2-}$ ionic diameter (4.42~\r{A}) and mass were used to define reduced units.
All interatomic energy scales were set to $\epsilon = 295 k_B$.
Time, temperature, and pressure were expressed in reduced units defined in the main text.

\subsection*{Ewald summation parameters}

Long-range electrostatic interactions were computed using the Ewald summation method.
The Ewald parameter was set to $\alpha = 0.6$,
the real-space cutoff was $6 d_{\rm Te}$,
and the reciprocal-space cutoff was $11 (2\pi/L)$.
These choices ensured a root-mean-square force error below $10^{-4}$.

\subsection*{Bond-orientational order parameters}

Local bond-orientational order parameters were defined as
\[
\bar{Q}^i_{lm} = \langle Y_{lm}(\hat{r}_{ij}) \rangle,
\]
\[
\hat{W}^i_l =
\frac{W^i_l}{\left( \sum_{m=-l}^{l} |\bar{Q}^i_{lm}|^2 \right)^{3/2}},
\]
where $Y_{lm}$ are spherical harmonics and $W^i_l$ involves Wigner 3$j$ symbols following standard definitions.

For the SC criterion, particles with coordination number six and
$0.14 < \hat{W}_4 < 0.16$ were classified as crystalline.
For the FCC-based criterion applied to Te$^{2-}$,
thresholds with $-0.085 < \hat W_4 < -0.169, -0.0109 < \hat W_6 < -0.0193, -0.0180 < \hat W_8 < 0.0640$
were adopted following Ref.~\cite{Kurita2010}.

\subsection*{Experimental details}
Polycrystalline AgPbBiTe$_3$ samples were synthesized following established procedures~\cite{sawahara2023}. 
Hard x-ray photoemission spectroscopy (HAXPES) measurements were performed at the Max-Planck-NSRRC HAXPES endstation~\cite{Takegami2019} at the Taiwan undulator beamline BL12XU of SPring-8 with an MB Scientific A-1 HE analyzer. The photon energy was set to 6.5~keV and the total energy resolution was about 250~meV. The sample was fractured under an ultrahigh vacuum of 10$^{-9}$~mbar in order to expose a fresh surface for the measurement. X-ray incidence angles of about $5^\circ$ and $45^\circ$, corresponding to emission angles of $85^\circ$ and $45^\circ$ respectively. The beam spot diameter was of about 50 $\mu$m. The measurements were performed at 300~K. Wide scans were performed in order to ensure the cleanliness of the measurement surfaces, and that no O nor C contaminants were present.

\section*{Supplementary results}
\subsection*{Crystallization dynamics}
We examined crystallization following a rapid quench from an equilibrated liquid at $T = 4.4$.
The melting temperature of AgPbBiTe$_3$ was determined to be $T_m = 2.62$.
Upon quenching to $T = 2.0$, the simulation box volume decreases with time, whereas at $T = 2.1$ the volume remains constant (Fig.~S1a).
The anion-anion radial distribution function $g(r)$ at $t = 2000$ shows sharp Bragg-like peaks at $T = 2.0$, confirming crystallization (Fig.~S1b).
Analogous simulations were performed for the charge-unified ideal (CUI) system.
Quenching below its melting temperature results in volume reduction and sharp $g(r)$ peaks at $T = 2.4$ (Fig.~S1c,d).
In both systems, the crystalline phase adopts the rocksalt structure.

\begin{figure}[htbp]
\centering
\includegraphics[width=14cm]{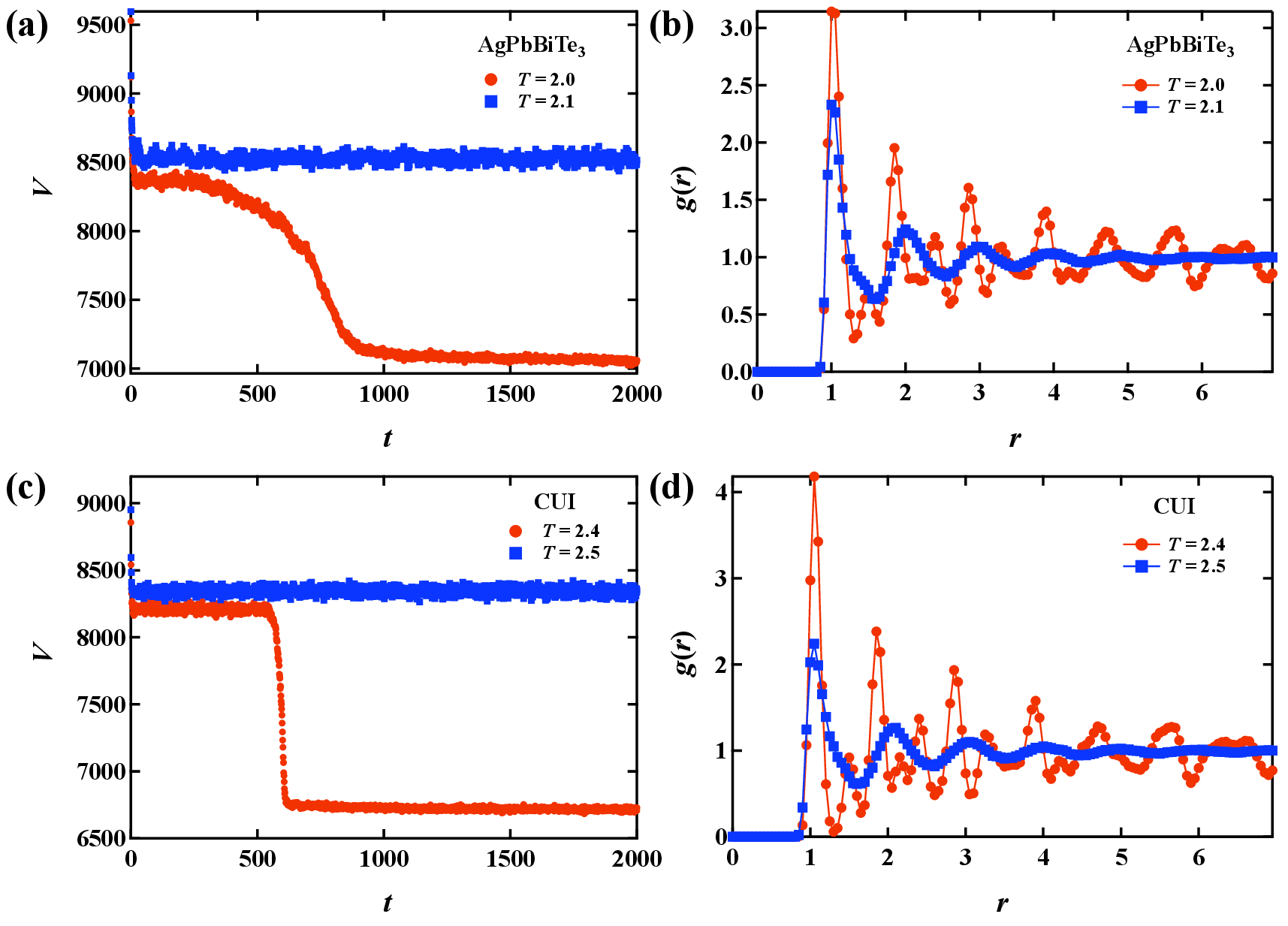}
\caption{
Crystallization dynamics.
(a) Volume evolution in AgPbBiTe$_3$ after quenching to $T = 2.0$ and 2.1.
(b) Corresponding anion-anion radial distribution functions.
(c,d) Volume evolution and $g(r)$ for the CUI system at $T = 2.4$.
Both systems crystallize into the rocksalt structure.
}
\label{crystal}
\end{figure}

\begin{figure}[htbp]
\centering
\includegraphics[width=14cm]{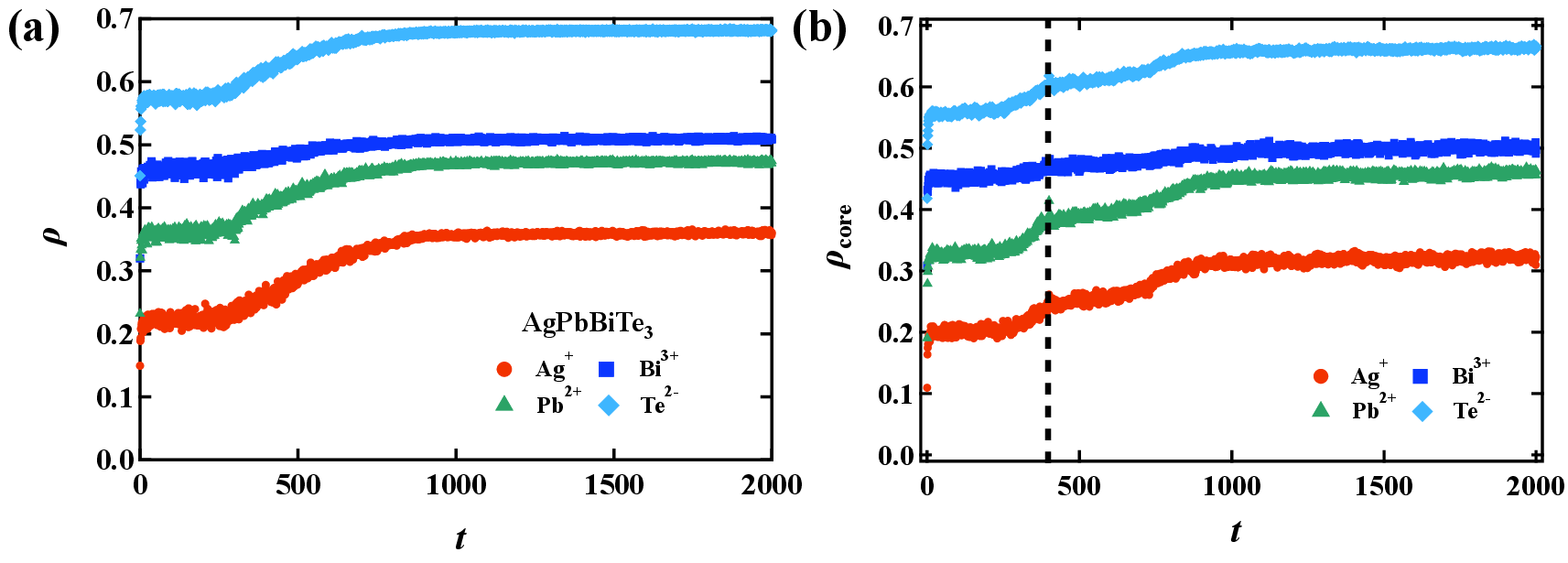}
\caption{\textbf{Time evolution of the average Voronoi density.}
(a) Time evolution of the average Voronoi density for each ion species in AgPbBiTe$_3$, respectively.
(b) Density evolution of particles classified as crystalline at $t$ = 400; the dotted line marks $t$ = 400. The density continues to increase even after crystallization.
}
\label{density}
\end{figure}

The local density of particle $i$ was computed from its Voronoi volume and ionic diameter.
Figure S2(a) show the average density of crystalline particles in AgPbBiTe$_3$.
The density increases gradually after crystallization begins, indicating that densification proceeds continuously during growth.
Classical nucleation theory would predict an abrupt density increase at the moment a particle becomes crystalline.
Instead, we observe a sequential process in which particles first acquire crystalline orientational order and subsequently densify.
Tracking particles identified as crystalline at $t = 400$ confirms that their density continues to increase over time (Fig.~\ref{density}(b)), demonstrating that densification predominantly occurs after structural ordering is established.

\subsection*{Interfacial Ag enrichment analysis}
To quantify cation composition near the crystal-liquid interface, we compare the numbers of crystalline particles identified by the SC and FCC criteria.
In the rocksalt structure, the SC criterion requires sixfold coordination and well-defined orientational symmetry.
Particles located at the outermost crystalline shell often lack complete coordination and are therefore not classified as crystalline under the SC criterion.
In contrast, the FCC-based criterion is applied to the Te$^{2-}$ sublattice.
Because nearest-neighbor cations surrounding crystalline Te sites are included in this definition, cations adjacent to the crystalline domain are counted even if they are located at the outermost shell.
As a result, the FCC criterion consistently identifies a larger number of cations than the SC criterion.
The difference $N_{\rm fcc} - N_{\rm sc}$ therefore provides an estimate of the number of cations located in the interfacial region adjacent to the ordered Te network.

Figure~S3 shows the time evolution of $N_{\rm fcc} - N_{\rm sc}$.
Within this interfacial population, the Ag fraction was calculated as $\frac{N_{\rm Ag}}{N_{\rm cation}}$, 
where $N_{\rm cation}$ denotes the total number of cations contributing to $N_{\rm fcc} - N_{\rm sc}$.
The resulting Ag fraction is approximately 0.5, significantly larger than the random expectation (1/3), demonstrating preferential Ag accumulation near the crystal interface.

\begin{figure}[htbp]
\centering
\includegraphics[width=14cm]{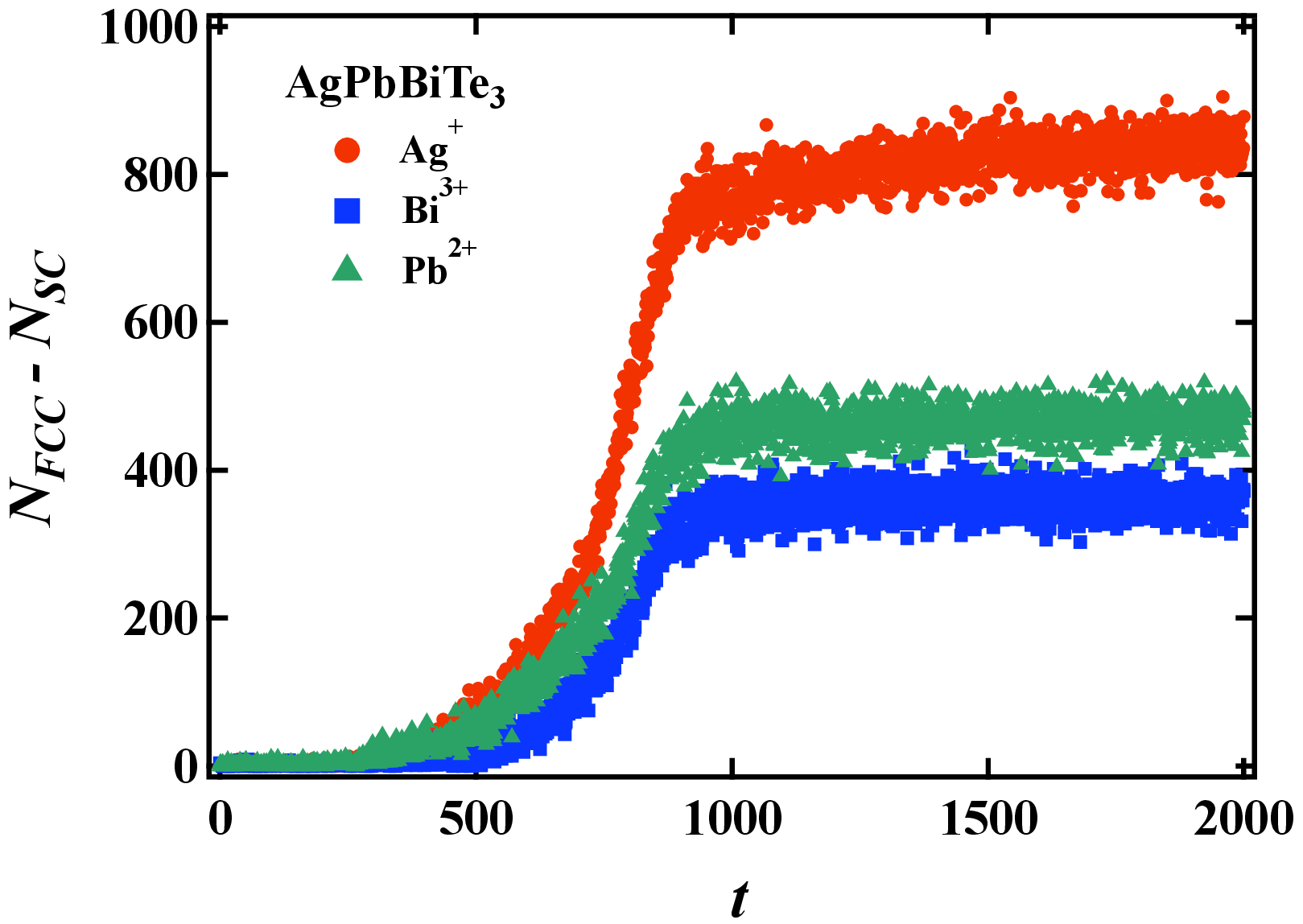}
\caption{$N_{\rm fcc} - N_{\rm sc}$ as a function of time. it represents the number of cations located in the interfacial region. 
This indicates that Ag ions are enriched in the interfacial region. 
}
\label{S-crystal}
\end{figure}

\subsection*{Depth-dependent photoemission measurements}
Polycrystalline AgPbBiTe$_3$ samples were mechanically fractured \textit{in-situ} prior to the photoemission measurement. In polycrystalline samples, the fracturing preferentially exposes grain-boundary surfaces, as grain boundaries are mechanically weaker regions.
It is expected, thus, that surface-sensitive measurements will be more representative of grain-boundary regions than bulk-sensitive measurements. By making use of hard x-ray photoemission spectroscopy, that is, photoemission spectroscopy using high energy photons, inelastic mean free paths of around $\lambda\sim50-100$\r{A} are obtained~\cite{Powell2016}, allowing for bulk-sensitive signatures to be measured. 
In order to decrease the effective probing depth, the emission angle $\theta$ can be set farther from normal emission. This will increase the distance a photoelectron needs to travel through the solid towards the analyzer, resulting in an effective scaling of the reachable depth corresponding to $\lambda \sin \theta$. Thus, smaller $\theta$ enhances surface sensitivity and increases the relative contribution from grain-boundary regions.

\begin{figure}[htbp]
\centering
\includegraphics[width=14cm]{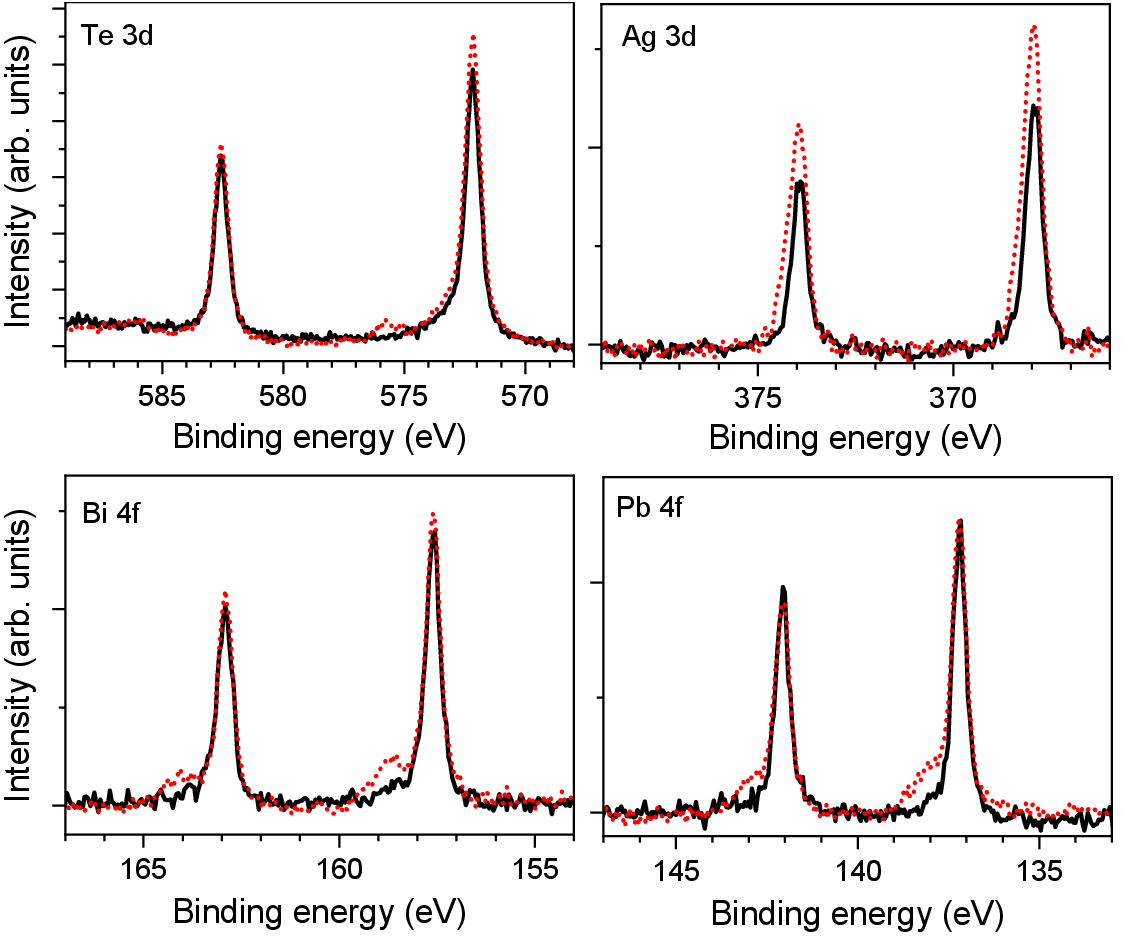}
\caption{
Photoemission core level spectra measured at bulk-sensitive (solid line) and surface-sensitive (dotted line), after a standard Shirley background subtraction and normalized to the total integrated intensity of the Te~$4d$ spectra. The measurements performed at $\theta = 45^\circ$ displays additional peaks and/or broader main peaks. We observe that the intensity of the Ag significantly increases for its surface-sensitive spectra, while the Bi and Pb only shows a modest change. This behaviour confirms a preferential enrichment of Ag in grain-boundary regions.
}
\label{XPS}
\end{figure}

Figure~\ref{XPS} shows photoemission core level spectra measured at $\theta = 85^\circ$ and $\theta = 45^\circ$, after a standard Shirley background subtraction and normalized to the total integrated intensity of the Te~$4d$ spectra. We observe that the bulk-sensitive spectra display, in all cases, a single sharp main peak for each core level spectrum, indicating a clear predominance of one single chemical component for each cation, as it would be expected from its predominantly bulk signal. Meanwhile, the more surface-sensitive spectra, obtained from the measurements performed at $\theta = 45^\circ$ displays additional peaks and/or broader main peaks. This confirms that this spectra is more representative of the surface vicinity, where ions with distinct chemical environments or characteristics than those from the bulk are present, resulting in these additional components. When comparing the $\theta = 85^\circ$ and $\theta = 45^\circ$ spectra, we observe that the intensity of the Ag significantly increases for its surface-sensitive spectra, while the Bi and Pb only shows a modest change. This behaviour confirms a preferential enrichment of Ag in grain-boundary regions.

\end{document}